%% file: main.tex
\documentclass[12pt]{article}

\usepackage[utf8]{inputenc}
\usepackage{amsmath, amssymb, amsthm, mathtools, bbm, graphicx, fullpage, caption, subcaption, multirow, url, authblk, hyperref}
\usepackage[dvipsnames]{xcolor}
\usepackage[normalem]{ulem}
\usepackage{rotating}
\usepackage{setspace}
\usepackage{booktabs}
\usepackage{booktabs}   
\usepackage{tabularx}   
\usepackage{amsmath}    
\usepackage{natbib}
\usepackage{lineno}

\title{Modeling Metabolic State Transitions in Obesity Using a Time-Varying Lambda–Omega Framework}

\author[1]{\small{Soheil Saghafi}}
\author[1,2]{\small{Gari D. Clifford}}
\affil[1]{\small{Department of Biomedical Informatics, Emory University}}
\affil[2]{\small{Department of Biomedical Engineering, Georgia Institute of Technology \& Emory University}}

\date{}

\begin{document}

\maketitle

%
%

\begin{abstract}
Obesity develops gradually over extended periods rather than emerging abruptly. Because this progression is slow and often subtle, physiological changes may remain unrecognized until excess adiposity is already established. Weight reduction is commonly framed as a simple consequence of “eating less and moving more”; however, this view overlooks the complex and adaptive nature of human metabolic regulation. During weight loss, resting metabolic rate often decreases and metabolic efficiency increases, making it progressively more difficult to maintain a caloric deficit. In contrast, during periods of overfeeding, increases in resting metabolic rate, thermogenesis, and spontaneous physical activity partially counteract excess energy intake and slow weight gain. Importantly, these compensatory responses are asymmetric, with stronger and more persistent adaptations occurring during underfeeding than during overfeeding. This asymmetry contributes to the gradual development of obesity and the biological difficulty of sustaining long-term weight loss. Here we introduce a dynamical-systems framework based on a lambda–omega model to describe metabolic regulation under lifestyle perturbations. By allowing the regulatory parameters to vary over time, the model captures progressive shifts in metabolic set-points and enables exploration of transitions between metabolic states underlying long-term weight gain and loss trajectories.
\end{abstract}

\section{Introduction}
Obesity is a chronic and multifactorial condition that develops gradually rather than emerging as a discrete event. Its progression is driven by long-term interactions between physiological, behavioral, and environmental factors \cite{muller2016energy, hill2012environment}, many of which unfold too slowly for individuals to recognize until substantial weight gain has already occurred. Historically, obesity was relatively rare worldwide; mid-20th-century populations exhibited markedly lower obesity prevalence due in large part to more physically demanding occupations, greater daily movement, and limited access to calorie-dense, ultra-processed foods \cite{church2011trends}. However, over the past five decades, rapid mechanization, shifts toward sedentary work, automobile-dependent urban design, and increased screen-based lifestyles have collectively reduced daily energy expenditure and contributed to what is now termed an \textit{obesogenic environment} \cite{swim2011environment}. Global surveillance data show that obesity has more than tripled since 1975 \cite{whofactsheet2024}, reflecting this profound societal transformation. Obesity is causally linked to a wide array of chronic diseases, including type~2 diabetes, cardiovascular disease, certain cancers, osteoarthritis, sleep apnea, and reproductive disorders \cite{guh2009incidence, lauby2016body}. Many of these conditions show significant improvement—or even complete remission—when obesity is effectively prevented or reversed. These historical and epidemiological patterns suggest that the modern obesity epidemic reflects sustained changes in environment and behavior rather than inevitable human biology. Evidence indicates that long-term restoration of habitual physical activity and healthier dietary patterns would substantially reduce population-level obesity risk for most individuals \cite{hall2011quantitative, swimburn2011obesogenic}.

Although weight management is often framed through the simplistic lens of reducing caloric intake and increasing physical activity, this perspective overlooks the complex regulatory systems that govern human metabolism and energy balance \cite{hall2012energy}. A large body of research has demonstrated that the body actively resists weight loss through adaptive changes in resting metabolic rate, hormonal signaling, appetite regulation, and energy expenditure. These compensatory processes—commonly referred to as \textit{\textbf{metabolic adaptation}} or \textit{\textbf{adaptive thermogenesis}}—contribute to weight-loss resistance \cite{rosenbaum2010adaptive, speakman2013adaptive} and help explain why many individuals struggle to maintain long-term reductions in body weight despite adherence to lifestyle interventions. Traditional conceptualizations of weight regulation, which assume linear relationships between diet, physical activity, and body mass, do not fully capture this dynamic and nonlinear physiological behavior \cite{hall2011biologically}. Despite advances in understanding metabolic adaptation, quantitative models that describe \textit{how} the body transitions between metabolic states—particularly in response to sustained lifestyle modifications—remain limited. Existing models of body-weight regulation often rely on static energy-balance equations or steady-state predictions, which assume rapid convergence to equilibrium and therefore provide limited insight into transient dynamics, stability, and long-term adaptation \cite{thomas2014mathematical, hall2011quantitative, chow2016dynamic}. Such approaches are inherently limited in their ability to capture nonlinear feedback, oscillatory behavior, hysteresis, and history-dependent responses that are fundamental features of metabolic regulation \cite{muller2016energy, kevinhall2012metabolic, speakman2014evolutionary}. These limitations motivate the development of mathematical frameworks that explicitly model metabolic regulation as a nonlinear, dynamical system evolving over time rather than as a static balance of energy fluxes.

To address this gap, the present study introduces a dynamical-systems approach to modeling metabolic adaptation using the lambda--omega ($\lambda$--$\omega$) framework. In contrast to classical formulations that assume fixed regulatory parameters, we introduce time-varying coefficients within the $\lambda$--$\omega$ system, allowing the governing parameters to evolve gradually in response to sustained environmental and physiological stressors. This formulation enables the model to represent the progressive drift of metabolic regulation over time rather than assuming a static attractor structure.

Importantly, this framework provides a mathematical bridge between the long-debated biological “set-point” theory—emphasizing genetically influenced homeostatic regulation—and the environmental “settling-point” theory—emphasizing lifestyle and environmental drivers of body weight. In our formulation, the baseline limit cycle reflects intrinsic biological regulatory tendencies (the set-point), while slow time-dependent modulation of $\lambda(t)$ and $\omega(t)$ captures environmentally induced shifts in metabolic stability (the settling-point). Rather than treating these perspectives as mutually exclusive, the model unifies them: genetic and physiological constraints define the system’s initial stability structure, whereas chronic environmental exposure reshapes that structure dynamically over time.

A key motivation for this framework arises from the asymmetric metabolic responses observed during underfeeding versus overfeeding. During caloric restriction, the body exhibits a pronounced adaptive thermogenic response in which resting metabolic rate decreases, sympathetic activity is reduced, thyroid hormone signaling is suppressed, and appetite-regulating hormones shift to promote weight regain \cite{rosenbaum2010adaptive, hall2011biologically, dulloo1997adaptive}. These coordinated physiological changes produce a steep downward shift in metabolic expenditure, often exceeding what would be predicted solely from the loss of body mass. In contrast, chronic overfeeding induces a markedly weaker and slower compensatory response. Although small increases in thermogenesis do occur—sometimes referred to as ``luxuskonsumption''—they are typically insufficient to neutralize sustained caloric surpluses \cite{ravussin1985variations, leibel1995changes, speakman2013adaptive}. Metabolic rate rises modestly with weight gain, but the adaptive component is minimal and unfolds over long timescales. This asymmetry creates a fundamental energetic imbalance in which the body vigorously defends against weight loss while offering comparatively little protection against gradual weight gain. 

Within the proposed time-varying $\lambda$--$\omega$ formulation, even small but persistent caloric excesses gradually modify the system parameters, leading to slow deformation of the original limit cycle and the emergence of a new metabolic equilibrium. In this way, chronic environmental exposure does not merely displace the system—it reshapes the underlying dynamical landscape itself.

The slow progression of resting metabolic rate in response to chronic overfeeding is therefore a central driver of obesity development. By analyzing these transitions within the $\lambda$--$\omega$ model, we aim to quantify the time required for the body to adapt to new lifestyle conditions, identify the stability of the resulting metabolic states, and examine how gradual changes in metabolic rate shape long-term weight trajectories. Overall, this work contributes a novel mathematical perspective on obesity progression and reversal, demonstrating how the $\lambda$--$\omega$ model can be used to capture the nonlinear metabolic responses that underlie both weight-loss resistance and the slow, cumulative effect of overfeeding. By explicitly allowing regulatory parameters to evolve over time, the framework reconciles biological predisposition with environmental influence, offering a unified dynamical interpretation of how inherited metabolic regulation and lifestyle-driven perturbations jointly determine long-term body-weight trajectories.

\section*{Materials and methods}
\begin{figure}
\centering
\includegraphics[width=1.0\textwidth]{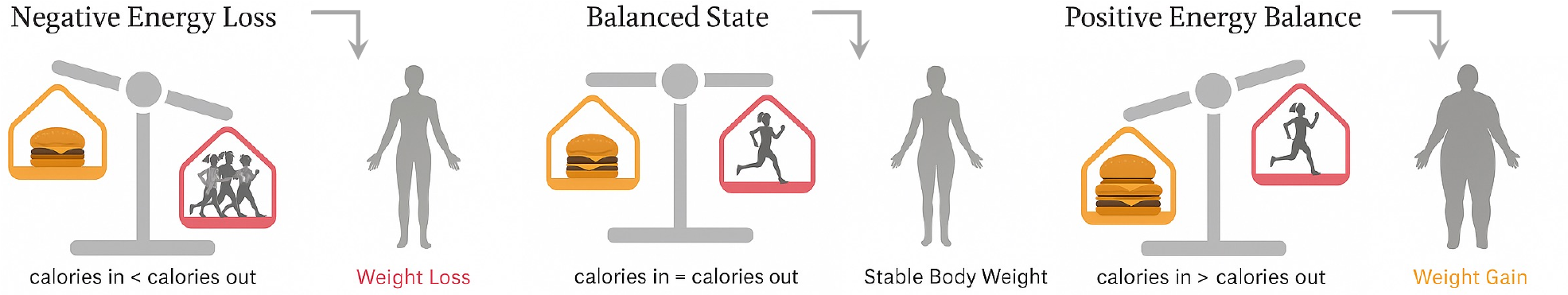}
\caption{\small Conceptual illustration of energy balance demonstrating how the relationship between caloric intake and energy expenditure governs changes in body weight. From left to right, the panels depict a negative energy balance, in which caloric intake is less than energy expenditure, resulting in weight loss; a balanced state, in which calories consumed equal calories expended and body weight is maintained; and a positive energy balance, where caloric intake exceeds energy expenditure and leads to weight gain.
}
\label{fig:Energy-Balance-Equilibrium}
\end{figure}

Body-weight regulation is fundamentally governed by energy balance, the dynamic interplay between caloric intake (``energy in'') and total energy expenditure (``energy out''). When energy intake and energy expenditure are balanced, body weight remains stable over time. Sustained negative energy balance, in which energy expenditure consistently exceeds caloric intake, leads to progressive weight loss as endogenous energy stores are mobilized. In contrast, sustained positive energy balance, characterized by caloric intake exceeding energy expenditure, results in weight gain through the gradual accumulation of body mass (see Fig. \ref{fig:Energy-Balance-Equilibrium}). Although this framework is conceptually simple, the underlying physiology is highly adaptive rather than static. As energy intake or body weight changes, the body actively modifies its energy expenditure, altering the very relationship that governs weight change over time.

This adaptive process---known as metabolic adaptation---describes the body's ability to adjust its energy requirements downward or upward in response to caloric restriction, overfeeding, weight change, or environmental demands. A central component of this adaptation is the basal metabolic rate (BMR), the energy required at rest to sustain essential physiological functions. BMR is shaped by multiple interconnected factors: body composition, with lean mass substantially elevating resting energy needs; metabolically active organs such as the liver, brain, heart, and kidneys, which collectively dominate baseline energy use; hormonal regulators including thyroid hormones, cortisol, growth hormone, and catecholamines that influence cellular metabolic activity; and aging, which reduces BMR through declining lean mass and diminished organ efficiency. Additional influences include overall body size, which affects thermoregulation and maintenance costs; genetic variation in organ size, muscle metabolism, and metabolic efficiency; environmental and physiological stressors such as illness or temperature extremes that can acutely raise energy expenditure; nutritional state, whereby caloric restriction suppresses BMR while overfeeding slightly increases it; and physical training, particularly resistance exercise, which increases muscle mass and elevates BMR. Together, these factors establish the baseline energetic state from which adaptation occurs. During weight loss, reductions in body mass, shifts in hormonal signals, and cues related to energy scarcity often produce a disproportionate decline in energy expenditure, resulting in a slower-than-expected metabolism that resists further weight loss. Conversely, during periods of overfeeding, modest increases in thermogenesis and metabolic activity can partially offset the energy surplus. By altering the energy-balance equation itself, metabolic adaptation helps explain why weight change is nonlinear, why individuals differ in their responses to dieting or overfeeding, and why lost weight is often challenging to maintain over time.

\begin{figure}
\centering
\includegraphics[width=1.0\textwidth]{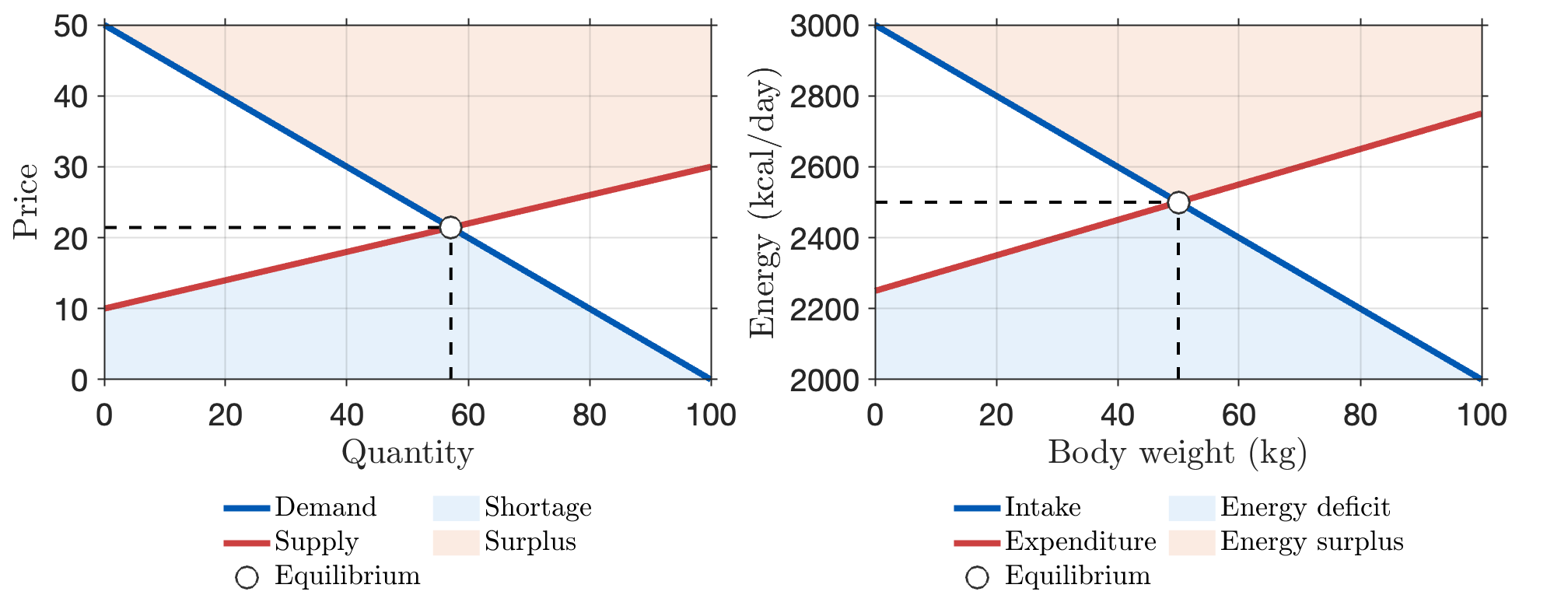}
\caption{\small Comparison between economic supply–demand equilibrium and physiological energy-balance equilibrium. Left: Classic supply–demand curves, where downward-sloping demand and upward-sloping supply intersect at a stable market equilibrium.
Right: Analogous intake–expenditure curves, transformed onto a regulatory axis to mirror the economic structure. Intake (demand-like) and expenditure (supply-like) intersect at the point of energy balance, representing stable body weight. Together, the panels illustrate that both markets and body weight regulation operate as self-balancing systems in which equilibrium is determined by the intersection of opposing input and output forces.}
\label{fig:Supply Demand Curve}
\end{figure}

The economic principle of supply and demand offers a powerful analogy for understanding how the body regulates energy and body weight, particularly through the concept of equilibrium. In a market, demand reflects how much consumers want a product, supply represents how much producers provide, and the equilibrium price emerges where the two forces balance. If demand exceeds supply, prices rise; if supply exceeds demand, prices fall. Similarly, the human body functions as a dynamic energy-balancing system in which energy intake (food consumption) acts like demand and energy expenditure acts like supply. The body’s weight represents the equilibrium point, adjusting up or down until intake and expenditure naturally match. When energy intake chronically exceeds expenditure, the body stores the surplus as fat, causing weight to rise until a new, higher equilibrium is reached. Conversely, when intake falls below expenditure, the body draws on stored energy, and weight declines until a new, lower equilibrium stabilizes. Just as markets adjust production and pricing in response to changing conditions, the body adapts its metabolic rate, appetite, and energy use as weight changes, constantly working toward restoring energy equilibrium. This perspective underscores that body-weight regulation is not a simple linear equation but a self-regulating equilibrium system, helping explain why sustained weight change is challenging and why obesity often emerges when long-term disruptions shift the body toward a higher energy-balance equilibrium. Figure \ref{fig:Supply Demand Curve} illustrates the conceptual parallel between economic market equilibrium and physiological energy balance. In the left panel, the supply–demand diagram illustrates a downward-sloping demand curve and an upward-sloping supply curve whose intersection marks the market equilibrium. At this point, the equilibrium price and quantity align so that supply equals demand, while the shaded regions highlight how deviations above or below this point generate surpluses or shortages, respectively. The right panel depicts an analogous relationship between energy intake and energy expenditure. To highlight the structural similarity, body‐weight–related regulatory signals were transformed onto a price-like axis, resulting in intake taking a downward (demand-like) slope and expenditure taking an upward (supply-like) slope. Their intersection represents the point of energy balance—where caloric intake equals caloric expenditure—which corresponds to a stable body weight. By placing these models side by side, the figure emphasizes that body-weight regulation, like a market system, is governed by opposing forces whose intersection defines a natural equilibrium. Shifts in appetite, metabolism, or environmental factors move the curves and therefore establish a new equilibrium, directly analogous to shifts in supply or demand in economic systems.

\subsection*{Mathematical model}

The $\lambda$--$\omega$ formalism originates in the work of Greenberg on oscillatory reaction--diffusion kinetics \cite{Greenberg1976,Greenberg1980}, where it served as a tractable amplitude--phase reduction for spatially extended chemical oscillators.  

It was subsequently adopted in the mathematical biology literature as a canonical model for limit-cycle oscillations and wave propagation (e.g., Murray \cite{Murray2002}, Lederman \cite{lederman2022parameter}). Further theoretical developments, such as those by Sherratt \cite{Sherratt1994}, explored the formation of wave trains and spatial patterns arising from $\lambda$--$\omega$ kinetics, while more recent work
(e.g., Flach \cite{Flach2007}) has continued to use this formulation as a baseline model for nonlinear oscillatory phenomena.  

In the present context, the classical autonomous $\lambda$--$\omega$ system provides a clean, analytically transparent setting in which to visualize nullclines, vector fields, amplitude relaxation, and peak-to-peak behavior associated with a stable limit-cycle oscillator.\\

\subsubsection*{A Lambda–Omega dynamical model of weight reduction with smoothly transitioning nullclines}

The classical autonomous $\lambda$--$\omega$ oscillator is a planar dynamical system that provides a canonical normal-form representation of oscillatory behavior near a supercritical Hopf bifurcation.  In Cartesian coordinates
$(x,y)$ with $r=\sqrt{x^{2}+y^{2}}$, the dynamics are

\begin{align}
    \dot{x} &= \lambda(r)\,x - \omega(r)\,y, \\
    \dot{y} &= \omega(r)\,x + \lambda(r)\,y,
\end{align}

where the real-valued function $\lambda(r)$ governs the radial (amplitude) evolution and $\omega(r)$ determines the angular frequency.  In the classical formulation we take

\begin{align}
    \lambda(r) &= 1 - r^{2}, \\
    \omega(r)  &= 2\pi,
\end{align}

which yields the well-known unit-radius limit cycle. Because $\lambda(r)>0$ for $0<r<1$ and $\lambda(r)<0$ for $r>1$, trajectories
spiral outward when inside the unit circle and inward when outside it, converging monotonically to the stable periodic orbit $r=1$. This system thus separates the amplitude dynamics (regulated entirely by $\lambda(r)$) from the phase dynamics (governed by $\omega(r)$), providing a simple analytic framework for studying nonlinear limit-cycle oscillations. To capture the transient dynamical changes that characterize metabolic adaptation, we model the system using a classical $\lambda$--$\omega$ oscillator with a time-dependent stability structure. The state of the system is represented in complex form as

\begin{equation}
  \dot{z}(t)
  = \bigl(\lambda(r,t) + i\,\omega(r)\bigr)\,z(t),
  \qquad
  z(t) = x(t) + i\,y(t),
\end{equation}

where $r = |z| = \sqrt{x^2 + y^2}$ denotes the oscillation amplitude. The function $\lambda(r,t)$ governs radial growth or decay, while $\omega(r)$ determines angular rotation. Transitioning to polar coordinates, $z = r e^{i\theta}$ yields

\begin{equation}
  \dot{r} = r\,\lambda(r,t),
  \qquad
  \dot{\theta} = \omega(r),
  \label{eq:lambda-omega-polar}
\end{equation}

so that the sign of $\lambda(r,t)$ determines whether oscillations grow, decay, or remain neutral, while $\omega(r)$ sets the approximate oscillation frequency. In Cartesian coordinates, the system becomes

\begin{equation}
\begin{aligned}
  \dot{x} &= \lambda(r,t)\,x - \omega(r)\,y,\\[2pt]
  \dot{y} &= \omega(r)\,x + \lambda(r,t)\,y,
\end{aligned}
\label{eq:lambda-omega-cartesian}
\end{equation}

which are the equations used for all numerical simulations and visualizations. To model oscillations with a well-defined amplitude, we adopt the standard quadratic form for the autonomous radial growth profile,

\begin{equation}
  \lambda_{\text{base}}(r)
  = \lambda_{\mathrm{scale}}\left(1 - r^{2}\right),
  \label{eq:lambda-base}
\end{equation}

which possesses a unique radial equilibrium at $r = 1$. When the overall sign is positive, $\lambda_{\text{base}}(r) > 0$ for $0 < r < 1$ and $\lambda_{\text{base}}(r) < 0$ for $r > 1$, making $r = 1$ a stable limit cycle. The constant $\lambda_{\mathrm{scale}}$ controls the rate at which trajectories converge to or diverge from this limit cycle. For simplicity, we take the angular frequency to be radius-independent,

\begin{equation}
  \omega(r) = \omega_{\mathrm{scale}}\,2\pi,
\end{equation}

so that trajectories rotate at an approximately constant angular speed. This choice isolates the effect of stability changes from any concurrent frequency variation. A central component of this study is a single smooth transition in stability, implemented through a time-dependent modulation of the radial growth term. Specifically, we define a scalar factor $s(t)$ that transitions
monotonically from $+1$ to $-1$ over a prescribed time window $[T_{\mathrm{start}},\,T_{\mathrm{end}}]$ using a cosine profile,

\begin{equation}
  s(t)
  =
  \begin{cases}
    +1, & t < T_{\mathrm{start}},\\[6pt]
    \cos\!\bigl(\pi (t - T_{\mathrm{start}})/T_{\mathrm{dur}}\bigr),
    & T_{\mathrm{start}} \le t \le T_{\mathrm{end}},\\[6pt]
    -1, & t > T_{\mathrm{end}},
  \end{cases}
\end{equation}

where $T_{\mathrm{dur}} = T_{\mathrm{end}} - T_{\mathrm{start}}$. The full time-dependent radial law is then

\begin{equation}
  \lambda(r,t) = s(t)\,\lambda_{\text{base}}(r),
\end{equation}

which flips the direction of radial flow. Before the transition, the origin is an unstable spiral and $r=1$ is a stable limit cycle; after the transition, the origin becomes a stable spiral and $r=1$ becomes unstable. During the transition both the vector field and the associated nullclines deform smoothly, producing a gradual, rather than abrupt, change in stability. This continuous deformation is central for visualizing how the system resists changes in observable oscillatory amplitude despite underlying structural shifts. In the planar $(x,y)$ representation, nullclines are the curves along which one component of the vector field has zero instantaneous change. The $x$-nullcline identifies points where $\dot{x}=0$, and the $y$-nullcline those where $\dot{y}=0$. Their intersections mark equilibrium points, and their orientations provide geometric insight into whether nearby trajectories spiral, stretch, or contract. Applying this definition to~\eqref{eq:lambda-omega-cartesian}, the nullclines satisfy

\begin{equation}
\begin{aligned}
  \lambda(r,t)\,x - \omega(r)\,y &= 0, \qquad (\dot{x}=0),\\[4pt]
  \omega(r)\,x + \lambda(r,t)\,y &= 0, \qquad (\dot{y}=0).
\end{aligned}
\end{equation}

Because both expressions depend explicitly on the time-varying term $\lambda(r,t)$, the nullclines rotate and deform smoothly as the stability signal $s(t)$ transitions from $+1$ to $-1$. Geometrically, this continuous rearrangement reflects the system’s shift from outward-spiraling to inward-spiraling trajectories. The left panel of the figure illustrates how the evolving nullcline geometry shapes the flow over time.

\begin{table}[ht]
\centering
\caption{Core energetic quantities and time interpretation in the $\lambda$--$\omega$ framework.}
\begin{tabularx}{\linewidth}{@{}l l X@{}}
\toprule
\textbf{Quantity} & \textbf{Unit} & \textbf{Meaning / Time Interpretation} \\
\midrule
Resting Metabolic Rate (RMR) 
& kcal/day 
& Baseline resting energy expenditure rate (reported per 24 hours). \\

Metabolic Adaptation ($\Delta \mathrm{RMR}$) 
& kcal/day 
& Suppression of measured RMR relative to predicted RMR. 
The magnitude is expressed in kcal/day, whereas its persistence is evaluated over weeks, months, or years \cite{maclean2011biology, sumithran2011long}). \\

Angular Frequency ($\omega$) 
& 1/time 
& Behavioral oscillation frequency defined as $\omega = 2\pi/T$, where $T$ denotes the characteristic behavioral cycle length (e.g., weekly routines or dieting–relapse cycles over months to years). \\
\bottomrule
\end{tabularx}
\label{tab:energetic_quantities_time}
\end{table}

As summarized in Table~\ref{tab:energetic_quantities_time}, RMR and metabolic adaptation are expressed in units of kcal/day, consistent with their definition as instantaneous rates of energy expenditure standardized over a 24-hour interval. While the magnitude of metabolic adaptation is therefore quantified in kcal/day, its persistence unfolds over substantially longer temporal horizons, including weeks, months, and even years. Longitudinal evidence demonstrates that metabolic adaptation may remain significantly suppressed for years following intensive weight loss \cite{fothergill2016persistent}, underscoring the distinction between the daily energetic magnitude of adaptation and its long-term temporal persistence. By contrast, the parameter $\omega$ denotes an angular frequency with units of inverse time and is therefore explicitly tied to the temporal scale of the governing dynamical system. In the present framework, $\omega$ does not represent circadian (24-hour) oscillations, which largely average out in long-term body mass dynamics. Rather, $\omega = 2\pi/T$ characterizes slower behavioral oscillations that influence cumulative energy balance, where $T$ denotes the dominant period of recurrent lifestyle variation. Such cycles may correspond to weekly behavioral patterns, seasonal fluctuations, or longer dieting–relapse dynamics occurring over months to years. Accordingly, whereas RMR and metabolic adaptation describe energetic rates, $\omega$ parameterizes the temporal structure of behavioral forcing acting upon the weight-regulation system.

\subsubsection*{Second-order Lambda–Omega dynamical model of overfeeding with time-varying stability and a growing limit cycle}

We study a planar dynamical system written in complex form as
\begin{equation}
    \dot{z}(t)
      = \left( \Lambda(r,t) + i\,\Omega(r,t) \right) z(t),
      \qquad
      z(t) = x(t) + i\,y(t),
\end{equation}
where 
\[
    r(t) = |z(t)| = \sqrt{x(t)^{2}+y(t)^{2}}
\]
denotes the amplitude.  
The real-valued function $\Lambda(r,t)$ governs radial growth or decay, while $\Omega(r,t)$ determines instantaneous angular velocity.  
Expressing $z = r e^{i\theta}$ yields the polar-coordinate system
\begin{equation}
    \dot{r}(t) = r(t)\,\Lambda(r,t),
    \qquad 
    \dot{\theta}(t) = \Omega(r,t).
    \label{eq:polar-system}
\end{equation}

Second order lambda-omega model assumes a quadratic radial dependence in both angular and radial components:
\begin{align}
    \Lambda(r,t) &= \lambda(t) - b r^{2}, 
    \label{eq:lambda}\\[3pt]
    \Omega(r,t) &= \omega_{0} + a r^{2},
    \label{eq:omega}
\end{align}
with constants $a,b>0$.  
In Cartesian coordinates, the system becomes
\begin{equation}
\begin{aligned}
    \dot{x} &= \Lambda(r,t)\,x - \Omega(r,t)\,y,\\
    \dot{y} &= \Omega(r,t)\,x + \Lambda(r,t)\,y,
\end{aligned}
\label{eq:cartesian-system}
\end{equation}
where $r=\sqrt{x^{2}+y^{2}}$.

To model a gradual physiological or metabolic transition, the radius of the system’s oscillatory behavior is allowed to evolve in time.  
We prescribe a time-varying target radius
\begin{equation}
    R_{\mathrm{target}}(t)
    = R_{\max}
      - \bigl(R_{\max}-R_{\min}\bigr)e^{-t/\tau_{\mathrm{grow}}},
    \label{eq:target-radius}
\end{equation}
where $R_{\min}$ is the initial oscillation amplitude, $R_{\max}$ is the asymptotic amplitude, and $\tau_{\mathrm{grow}}$ defines the timescale of adaptation.  

To ensure that the radial dynamics stabilize at $r(t)\approx R_{\mathrm{target}}(t)$, we choose
\begin{equation}
    \lambda(t) = b\,R_{\mathrm{target}}(t)^{2}.
    \label{eq:lambda-of-time}
\end{equation}
Substituting \eqref{eq:lambda-of-time} into \eqref{eq:polar-system} yields the instantaneous radial equilibrium condition
\begin{equation}
    \dot{r}=0
    \quad \Longrightarrow \quad
    r(t) \approx R_{\mathrm{target}}(t),
\end{equation}
so that the system tracks a slowly expanding limit cycle whose radius increases monotonically toward $R_{\max}$.

In a two-dimensional dynamical system, nullclines mark the curves along which one component of the vector field momentarily stops changing. The $x$-nullcline consists of all points where $\dot{x}=0$, and the $y$-nullcline consists of all points where $\dot{y}=0$. Their intersections indicate equilibrium points, while their orientations and relative positions provide geometric insight into how trajectories move, spiral, or contract in different regions of the phase plane.

For the system in~\eqref{eq:cartesian-system}, the nullclines are given by
\begin{align}
    \dot{x} = 0 &\iff \Lambda(r,t)x - \Omega(r,t)y = 0,\\
    \dot{y} = 0 &\iff \Omega(r,t)x + \Lambda(r,t)y = 0.
\end{align}
These expressions define two time-dependent curves whose geometry evolves continuously because both 
$\Lambda(r,t)$ and $\Omega(r,t)$ depend on the amplitude $r^{2}$ and the time-varying growth rate $\lambda(t)$. The sign of $\Lambda(r,t)$ governs radial stability. When
\[
\begin{aligned}
    &\Lambda(r,t) > 0 \;\Rightarrow\; \text{outward growth (unstable radial dynamics)} \qquad\\[6pt]
    &\Lambda(r,t) < 0 \;\Rightarrow\; \text{radial contraction} \qquad
\end{aligned}
\]

Thus the curve defined implicitly by $\Lambda(r,t)=0$ specifies the instantaneous radius of the limit cycle and demarcates the transition between expanding and contracting dynamics.

\subsubsection*{Linking overfeeding adaptation to weight-loss dynamics through a second-order Lambda–Omega model}

To examine how metabolic adaptation operates differently during weight gain and weight loss, we extended the classical second order $\lambda$–$\omega$ oscillator to allow for a time–dependent limit–cycle radius whose evolution reflects physiological changes in energetic efficiency. In the overfeeding scenario, the radius of the limit cycle grows over time, representing the progressive amplification of oscillatory excursions in energy intake, expenditure, and weight dynamics as the system defends a higher set-point.










To capture the fundamentally different physiological behavior during weight-loss interventions, we inverted the structure of the model by enforcing a shrinking limit–cycle radius. Empirically, weight loss is characterized by increasing metabolic efficiency, 
reduced variability in energy flux, and a tendency for the system to collapse toward a stable, lower–amplitude attractor. To represent this, we define

\begin{equation}
    R_{\text{target}}(t) 
    = R_{\min}
      + \bigl(R_{\max} - R_{\min}\bigr)
        e^{-t/\tau_{\text{shrink}}},
\end{equation}

with $R_{\max}$ denoting the initial metabolic amplitude and $R_{\min}$ the reduced oscillatory radius associated with post–adaptation weight maintenance. As $t \to \infty$, $R_{\text{target}}(t)$ converges monotonically to 
$R_{\min}$, mirroring the progressively reduced metabolic flexibility during continued caloric restriction or sustained physical activity. 




In this formulation, the oscillatory amplitude gradually diminishes from cycle to cycle, and the system responds more efficiently to perturbations as the trajectory moves inward. Ultimately, the dynamics settle onto a low-amplitude stable attractor that reflects the metabolically adapted state associated with the post--weight-loss regime.//

\subsubsection*{Code Availability and Simulation Framework}
To enable reproducibility and further exploration of the proposed dynamical framework, we developed a MATLAB-based simulator that implements the time-varying lambda--omega system described in this study. The nonlinear ordinary differential equations are numerically integrated using MATLAB’s adaptive Runge--Kutta solver \texttt{ode45}, with strict numerical tolerances to ensure accurate trajectory computation during time-varying parameter transitions. The simulator produces animated phase-plane visualizations, including the evolving vector field, nullclines, and system trajectory, along with a time-series panel summarizing peak dynamics of $x(t)$. The full implementation and scripts required to reproduce the simulations and figures are publicly available at: \textcolor{blue}{\url{https://github.com/cliffordlab/MetabolicStateTransitionsModel/tree/main}}.

\section*{Results}

\begin{figure}
\centering
\includegraphics[width=0.45\textwidth]{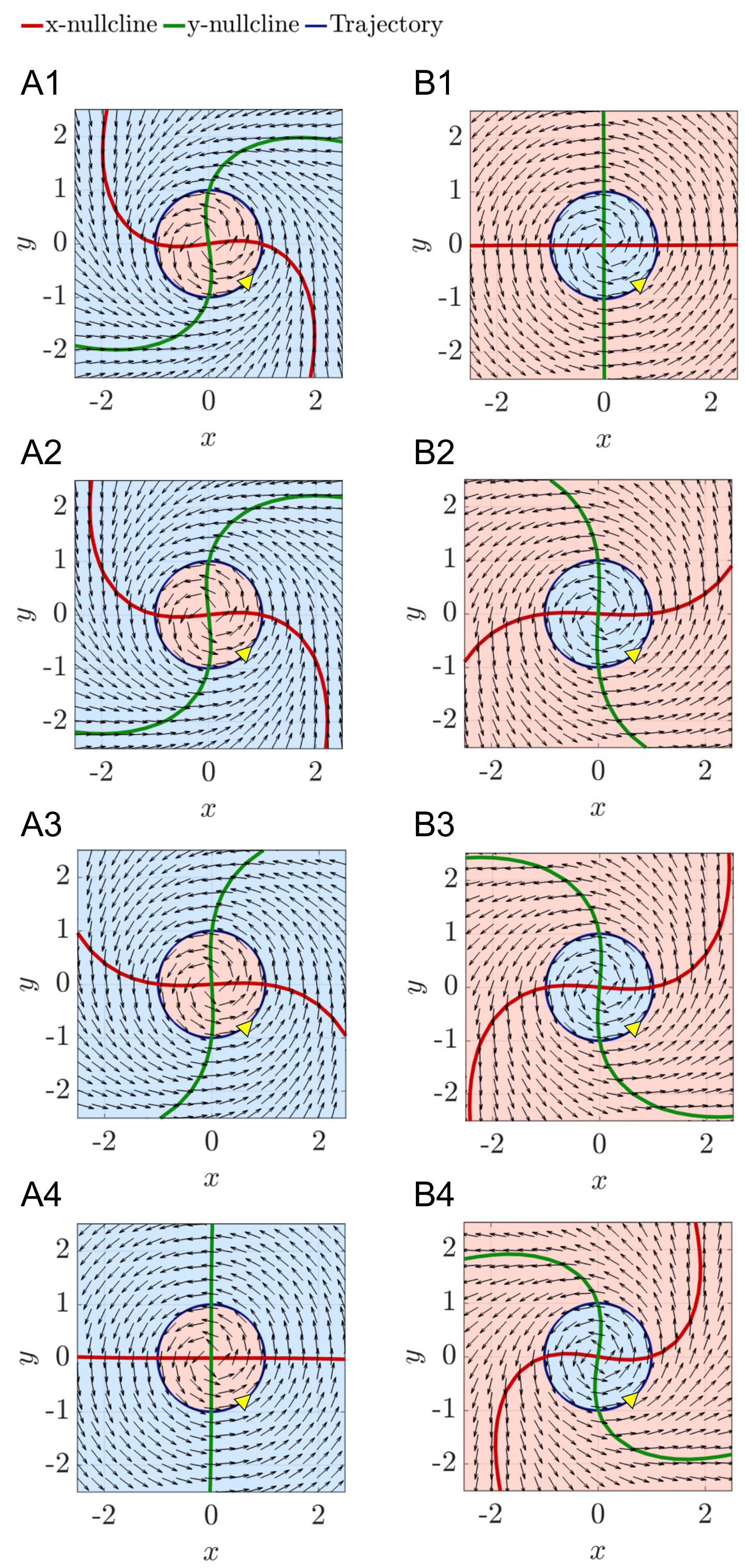}
\caption{\small \textbf{Smooth transition in nullcline geometry induces a change from unstable to stable spiral dynamics.}
Phase--plane snapshots of the $\lambda$--$\omega$ system illustrating how gradual deformation of the nullclines alters the stability of the fixed point. The left column (A1--A4) shows the initial configuration in which the nullclines intersect to produce an unstable spiral. As they shift smoothly from the top left to bottom right, the fixed point transitions through a bifurcation--like change in which instability is lost and stability emerges. The right column (B1–B4) depicts the subsequent strengthening of this stable state, with the fixed point evolving from weakly stable to strongly stable as the nullclines continue their smooth displacement. Red and blue shaded regions indicate the domains of outward growth and inward contraction, respectively, corresponding to instantaneous unstable and stable directions of the flow. Collectively, the panels demonstrate how continuous geometric changes in the nullclines can lead to abrupt qualitative shifts in system stability.}
\label{fig:phase-plane-limit-cycle}
\end{figure}

To illustrate how qualitative dynamics evolve as the nullclines undergo a smooth deformation, Figure~\ref{fig:phase-plane-limit-cycle} presents a two-column sequence of phase–plane snapshots from the $\lambda$–$\omega$ system. The left column (panels A1–A4) shows the configuration in which the nullclines initially intersect to form an unstable spiral fixed point. As the nullclines gradually transition—moving effectively from the top left toward the bottom right—the geometry of their intersection changes continuously, driving the system toward the regime in which the fixed point switches from unstable to stable. This smooth deformation of the nullclines therefore induces a bifurcation–like transition between instability and stability. The right column (panels B1–B4) then illustrates the continuation of this transition, beginning with a weakly stable fixed point and showing how the ongoing smooth shift of the nullclines progressively strengthens stability, ultimately yielding a strongly stable spiral that robustly attracts nearby trajectories. Across both rows, the red and blue shaded regions delineate the sectors of the phase plane in which perturbations respectively grow or decay, reflecting the instantaneous unstable and stable directions of the flow. Taken together, the panels illustrate that smooth, continuous shifts in the nullcline geometry can precipitate sharp qualitative transitions in stability and, once the system has crossed into the stable regime, progressively strengthen the attracting behavior of the fixed point.\\

As shown in Figure~\ref{fig:phase-plane-limit-cycle}, this transition from an unstable to a stable spiral provides a natural dynamical interpretation of metabolic adaptation. In the context of body-weight regulation, the initial unstable configuration represents the equilibrium state of an individual prior to sustained behavioral change. Perturbations such as overeating, caloric restriction, or increased physical activity temporarily displace the system, but the underlying regulatory mechanisms act to restore equilibrium. However, when these perturbations persist over time—such as during prolonged dieting or extended periods of high physical activity—the effective nullclines of the system begin to shift, gradually altering the geometry of the fixed point. Panels A1–A4 correspond to this early phase, in which the system still behaves as before despite the slow structural drift. Yet once a critical threshold is crossed, the type of the fixed point changes qualitatively, and the system transitions into the stable-spiral regime shown in panels B1–B4. This marks the onset of metabolic adaptation, during which metabolic rate decreases and the trajectories converge toward a new, strongly attracting equilibrium, producing the characteristic weight-loss plateau.

\subsubsection*{A Lambda–Omega dynamical model of weight reduction with smoothly transitioning nullclines}

\begin{figure}
\centering
\includegraphics[width=1.0\textwidth]{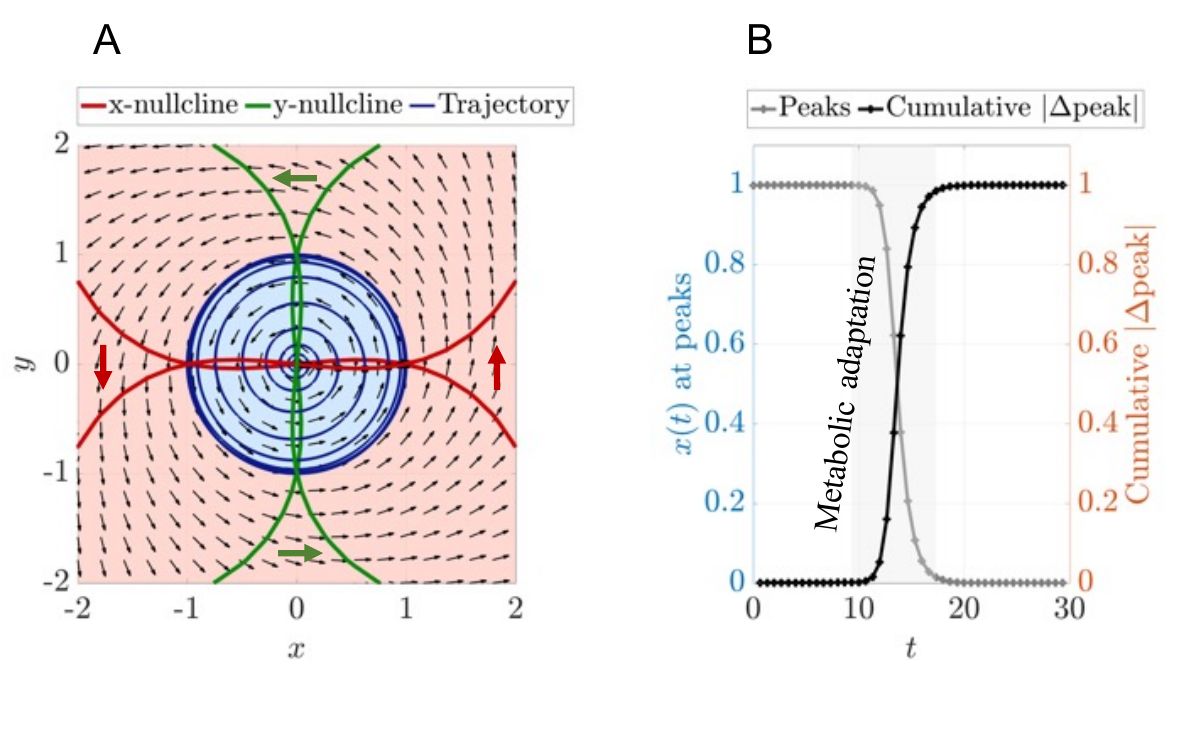}
\caption{\small \textbf{Dynamics of a time-varying $\lambda$--$\omega$ system illustrating the transition from an unstable to a stable fixed point and the emergence of metabolic adaptation.} 
\textit{Panel A:} The phase-plane shows a smooth and continuous transition of the nullclines, which gradually changes the stability structure of the system from an unstable fixed point (supporting sustained oscillations) to a stable fixed point (attracting trajectories inward). The trajectory initially exhibits outward spiraling behavior consistent with an unstable focus, and after the nullcline transition is completed, the dynamics reorient toward an inward spiral converging to the new stable equilibrium. 
\textit{Panel B:} The time series displays the peak amplitudes of the oscillations (gray; left axis) and their cumulative sum of these consecutive differences (black; right axis). During the early portion of the dynamics---when the fixed point is still unstable---peak amplitudes remain unchanged, indicating that the oscillatory regime persists despite the slow, ongoing structural transition. After several cycles following a perturbation, the system enters a new regime referred to as the \textit{metabolic adaptation phase} (shaded region). During this phase, the system resists changes in oscillation amplitude even as the underlying stability continues to evolve. Once the fixed point becomes fully stable, the peak amplitudes begin to decrease monotonically, signaling the attenuation of the adaptive response and the gradual convergence toward a plateau that represents the new metabolic steady state. Once the fixed point becomes fully stable, the peak amplitudes begin to decrease monotonically, marking the attenuation of the adaptive response and the approach to a plateau, representing convergence to the new metabolic steady state.}
\label{fig:metabolic-adaptation}
\end{figure}

To illustrate how a gradual change in system structure can generate the characteristic plateau-like behavior observed in metabolic adaptation, Figure~\ref{fig:metabolic-adaptation} presents the dynamics of a time-varying $\lambda$--$\omega$ model undergoing a smooth transition in its nullclines. As the nullclines shift, the fixed point of the system moves from an unstable to a stable configuration, and this continuous deformation produces a qualitative change in the trajectory geometry. Early in the transition, trajectories spiral outward, reflecting the presence of an unstable focus. Despite this ongoing structural drift, the oscillation amplitudes remain essentially constant, demonstrating a prolonged interval during which the system resists immediate change. This persistence constitutes the model’s analogue of a metabolic adaptation phase, in which the observable behavior is temporarily maintained even though the underlying stability continues to evolve.

As the nullcline shift completes and the fixed point becomes fully stable, the system ceases to support sustained oscillations. The trajectory reorients into an inward spiral, and the peak amplitudes begin to decline. The accompanying time-series plot highlights this shift, with peak values plateauing during the adaptation interval and subsequently decreasing monotonically once stability is restored. The cumulative sum of these consecutive peak's differences measure further emphasizes this departure from the adaptive regime, marking the system’s progressive convergence toward a new metabolic steady state. Together, these dynamics illustrate how a time-varying stability landscape can naturally give rise to the delayed behavioral change and eventual plateau characteristic of metabolic adaptation.

\subsubsection*{Second-order Lambda–Omega dynamical model of overfeeding with time-varying stability and a growing limit cycle}

During weight gain, metabolic adaptation does occur, but it is markedly slower and smaller in magnitude than the compensatory responses observed during weight loss. As body mass increases, resting energy expenditure, the thermic effect of food, and non-exercise activity thermogenesis tend to rise modestly, providing only partial opposition to sustained caloric surplus \cite{leibel1995changes, rosenbaum2010adaptive, speakman2013adaptive}. These increases emerge gradually over long timescales and are often insufficient to prevent continued fat accumulation when excess intake persists \cite{hall2012energy, hill2003obesity}. In contrast, weight loss elicits rapid and disproportionate metabolic adaptations, including pronounced reductions in resting metabolic rate, suppressed sympathetic and thyroid signaling, and hormonal changes that promote hunger and energy conservation \cite{leibel1995changes, rosenbaum2008effects, dulloo2012adaptive}. This asymmetry reflects a fundamental bias in human energy regulation, whereby the body strongly defends against negative energy balance but provides relatively weak and delayed resistance to positive energy balance, favoring gradual weight gain while rendering sustained weight loss physiologically difficult \cite{speakman2013adaptive, hall2018energy}.

\begin{figure}
\centering
\includegraphics[width=1.0\textwidth]{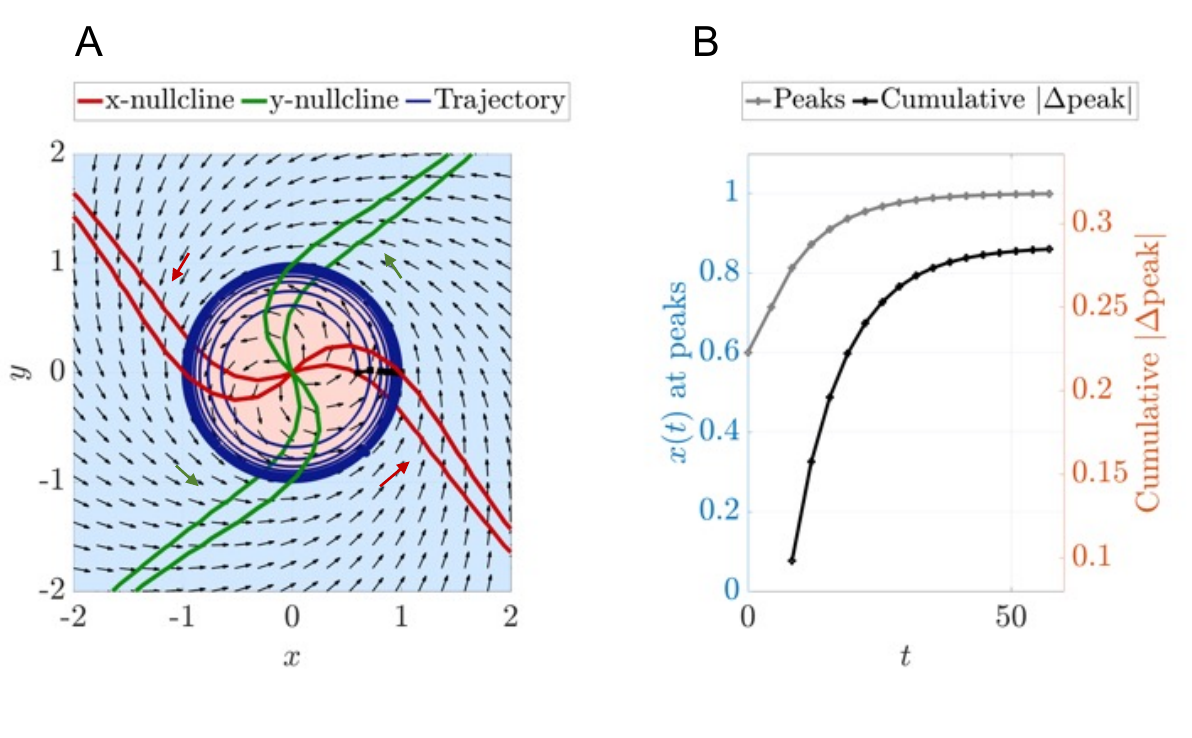}
\caption{\small \textbf{Dynamics of a time-varying second-order $\lambda$--$\omega$ system illustrating the smooth transition from a tiny limit cycle to the larger size limit cycle with unstable fixed point.} 
\textit{Panel A:} The phase-plane illustrates a smooth, continuous transition of the nullclines that gradually expands the size of the limit cycle. Regardless of whether the initial condition begins inside or outside the cycle, the trajectory is drawn toward it. As the nullclines shift outward and the limit cycle grows, the trajectory is correspondingly pushed in that direction, consistently guiding the system toward the evolving cycle.
\textit{Panel B:} This plot shows the change in peak amplitude from one cycle to the next (gray curve) and the cumulative sum of these consecutive differences, which reflects the accumulating effect of metabolic adaptation.}
\label{fig:overfeeding-case}
\end{figure}

This dynamical framework also suggests how to model the opposite scenario—chronic positive energy balance during overeating. As illustrated in Figure~\ref{fig:overfeeding-case}, sustained excess intake does not drive the system toward a more stable fixed point; rather, it causes the effective limit cycle to expand slowly over time. In this view, overeating is represented not by convergence toward a new equilibrium but by a gradual outward drift of the oscillatory radius. Such a mechanism also offers a conceptual strategy for weight-loss interventions, whereby keeping body-weight dynamics close to a slowly evolving limit cycle and guiding that limit cycle inward at a controlled pace may allow weight to decrease gradually without triggering abrupt metabolic adaptation. In the sections that follow, we formalize this hypothesis by introducing a model in which the limit-cycle radius evolves slowly under behavioral inputs, enabling us to compare the dynamical signatures of overeating, adaptation, and sustainable weight-loss trajectories within a unified mathematical framework.

\subsubsection*{Linking overfeeding adaptation to weight-loss dynamics through a second-order Lambda–Omega model}

\begin{figure}
\centering
\includegraphics[width=1.0\textwidth]{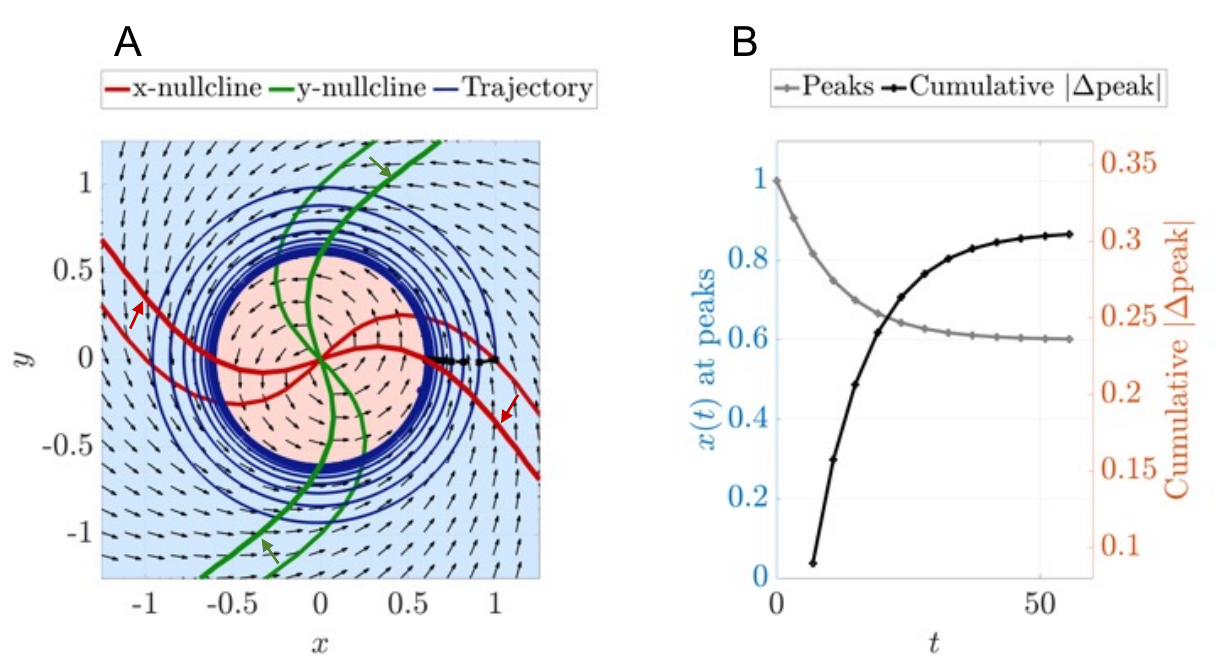}
\caption{\small \textbf{Dynamics of a time-varying second-order $\lambda$–$\omega$ system illustrating the smooth transition from a large limit cycle to a smaller one.} 
\textit{Panel A:} The phase-plane illustrates a smooth, continuous transition of the nullclines that gradually causes the limit cycle to shrink. Regardless of whether the initial condition begins inside or outside the cycle, the trajectory is drawn toward it. As the nullclines shift inward and the limit cycle contracts, the trajectory moves accordingly, consistently guiding the system toward the evolving cycle.
\textit{Panel B:} This plot shows the change in peak amplitude from one cycle to the next (gray curve) and the cumulative sum of these consecutive differences, which reflects the accumulating effect of metabolic adaptation.}
\label{fig:weight-loss-pattern-from-overfeeding}
\end{figure}

Figure~\ref{fig:weight-loss-pattern-from-overfeeding} illustrates how a gradual inward shift of the nullclines produces a slow contraction of the limit cycle, modeling the long-term metabolic adaptation that emerges when lifestyle changes—such as caloric restriction, increased physical activity, or sustained behavioral modification—are maintained consistently over time. Because the deformation of the nullclines is smooth rather than abrupt, the system transitions continuously toward a smaller metabolic steady-state, avoiding the sharp inflection that typically marks the onset of a weight-loss plateau during more aggressive dieting approaches. This behavior highlights a central insight of the model, namely that durable weight loss requires persistent, habitual adjustments that reshape the underlying dynamical landscape slowly enough for metabolic adaptation to follow without triggering compensatory resistance.

Interpreting this figure in the broader context of metabolic regulation also clarifies the fundamental asymmetry between weight gain and weight loss. Empirically, weight gain tends to occur gradually and often without immediate notice, whereas weight loss is more effortful and strongly constrained by adaptive reductions in energy expenditure. In our dynamical formulation, these opposing tendencies correspond to slow geometric deformations of the nullclines that reshape the underlying limit cycle. For the overfeeding scenario (Figure~\ref{fig:overfeeding-case}), the outward drift of the nullclines represents a metabolic environment in which the system is guided toward a larger-amplitude limit cycle, capturing the progressive increase in energy balance associated with sustained caloric surplus. Together, the two scenarios illustrate how smooth structural changes in the dynamical system can encode both the ease of weight gain and the resistance to weight loss through the lens of metabolic adaptation.

\section*{Discussion}

The present study introduces a dynamical-systems framework for understanding metabolic adaptation and weight-change trajectories, offering an interpretable alternative to traditional energy-balance models. Our findings highlight several conceptual and practical implications for obesity research, weight-loss interventions, and long-term metabolic regulation. A widely disseminated belief in public health messaging is that weight management can be reduced to the simple prescription of ``eating less and moving more.'' While caloric deficit remains a thermodynamic requirement for weight loss, decades of physiological research demonstrate that this reductionist view fails to account for the nonlinear and adaptive nature of human metabolism \cite{hall2011biologically, muller2016energy}. For example, caloric restriction reliably triggers compensatory reductions in resting metabolic rate, thyroid hormone activity, sympathetic nervous system output, and spontaneous physical activity, collectively reducing total energy expenditure beyond what is expected from changes in body composition alone \cite{rosenbaum2010adaptive, dulloo1997adaptive}. This phenomenon—commonly described as \textit{metabolic adaptation} or \textit{adaptive thermogenesis}—is one contributor to what clinicians refer to as \textit{metabolic confusion}, in which individuals adhering to a caloric deficit experience stagnation or reversal in weight loss despite consistent behavioral adherence.

Conversely, chronic overfeeding induces a much weaker adaptive response. While overfeeding can produce small increases in thermogenesis, these changes are insufficient to offset persistent energy surplus, thereby allowing gradual and often unnoticed weight gain to accumulate over years \cite{ravussin1985variations, leibel1995changes}. This asymmetric metabolic defense system provides a mechanistic explanation for why small, sustained caloric surpluses lead to obesity more reliably than caloric deficits lead to durable weight loss.

The weight-loss process is further complicated by psychological and behavioral factors. Numerous studies show that many individuals experience a decline in motivation and adherence once weight loss slows or reaches a plateau, even when beneficial metabolic changes continue beneath the surface \cite{teixeira2015successful}. Perceived failure during these plateaus can reduce self-efficacy, increase dropout rates, and promote compensatory eating behaviors, ultimately undermining long-term success \cite{linde2004measuring}. Importantly, plateaus are not indicative of metabolic failure but rather represent a period of physiological recalibration during which hormonal, lipid, and inflammatory markers continue to improve despite unchanged body weight \cite{franz2007weight, goodpaster2010effect}. Recognizing plateaus as a normal component of the adaptive process may improve psychological resilience and intervention retention.

Our dynamical-systems model emphasizes the importance of aligning the pace of weight loss with the pace of metabolic adaptation. Rapid weight loss can exaggerate adaptive thermogenesis, lowering energy expenditure sharply and thereby increasing the biological ``fight-back'' response that drives weight regain \cite{hall2012energy, rosenbaum2010adaptive}. In contrast, gradual and sustained weight loss allows metabolic rate to decline more slowly, producing a closer match between the trajectory of adipose-tissue reduction and the body's adaptive response. Such alignment minimizes abrupt decreases in resting metabolic rate and reduces compensatory appetite increases, improving the likelihood of long-term maintenance \cite{katan2009effect}. This interpretation aligns with empirical evidence showing that slower, behaviorally consistent weight loss is associated with better long-term outcomes than rapid dieting or highly restrictive interventions \cite{dombrowski2014interventions}.

Research consistently shows that \textit{metabolic adaptation is highly heterogeneous across individuals}, and this variability contributes substantially to the wide range of weight-loss responses observed under identical lifestyle interventions. Classic controlled feeding studies demonstrate that some individuals exhibit \textit{minimal reductions} in energy expenditure during caloric restriction, whereas others display \textit{adaptive thermogenesis} exceeding 300~kcal/day, far beyond what would be predicted from changes in body composition alone \citep{Leibel1995, Rosenbaum2010}. These differences arise from several sources, including \textit{genetic variation}, which accounts for a substantial proportion of the variance in metabolic responses \citep{Speakman2011}, as well as differences in baseline \textit{fat-free mass}, where individuals with lower muscle mass experience proportionally greater metabolic slowing \citep{Muller2016}. Hormonal responses also differ markedly—changes in leptin, thyroid hormones, and sympathetic outflow vary widely between individuals and strongly modulate both resting metabolic rate and appetite during weight loss \citep{Rosenbaum2008}. Behavioral compensation contributes additional variability, as reductions in spontaneous physical activity and non-exercise activity thermogenesis (NEAT) can differ by hundreds of kilocalories per day between individuals undergoing the same caloric deficit \citep{Levine1999}. Overall, these findings underscore the heterogeneity of metabolic adaptation, which manifests as a continuum of physiological strategies from pronounced energy conservation to attenuated compensatory responses. This inter-individual heterogeneity underscores why simple prescriptive advice to ``eat less and move more'' fails to reflect the \textit{dynamic and personalized} nature of human energy regulation. Taken together, our results illustrate that metabolic regulation is best understood as a nonlinear dynamical system characterized by feedback loops, asymmetric adaptive responses, and state-dependent stability properties. The lambda--omega framework provides an interpretable mathematical structure that captures these features and offers insight into why obesity develops gradually yet resists reversal. By modeling the transitions between metabolic states---rather than relying on static energy-balance assumptions---we highlight the importance of pacing, psychological resilience, and adaptive physiology in long-term weight management. This perspective may inform the design of interventions that explicitly account for metabolic adaptation, support behavioral persistence through plateaus, and better align physiological and behavioral trajectories over time.


Future work will focus on extending the proposed framework to account for inter-individual heterogeneity in metabolic responses to weight loss. Individuals differ substantially in biological background, physiological characteristics, and demographic factors, all of which can influence the magnitude, timing, and persistence of metabolic adaptation during weight loss. The current model provides a flexible foundation for incorporating additional parameters that capture these sources of variability. 

Importantly, longitudinal evidence demonstrates that early metabolic slowdown during intensive weight loss does not necessarily predict long-term weight regain; however, persistent metabolic adaptation years after weight loss is significantly associated with greater weight regain, suggesting a sustained biological defense of prior body weight \cite{fothergill2016persistent}. In a six-year follow-up of participants from \textit{The Biggest Loser} competition, metabolic adaptation remained suppressed relative to predicted resting metabolic rate even after substantial weight regain, and the magnitude of this persistent adaptation correlated with the degree of weight regained. These findings highlight that the temporal persistence of metabolic adaptation—rather than its acute magnitude—may be a critical determinant of long-term weight trajectories.

By systematically exploring meaningful parameter ranges, future studies can examine a spectrum of metabolic response scenarios and interpret these regimes in terms of underlying biological mechanisms. Such analyses may help link abstract model dynamics to measurable physiological traits, thereby offering insights into why weight-loss trajectories differ across individuals and informing the development of more personalized and effective weight-loss strategies.

\section*{Conclusion}

In this work, we introduced a mechanistic dynamical-systems framework based on a time-varying $\lambda$--$\omega$ model to characterize the nonlinear processes underlying metabolic adaptation during overfeeding and weight loss. By allowing 
the nullclines to shift smoothly over time, the model captures how the stability structure of the system can gradually transition from an unstable oscillatory regime to a stable equilibrium. This continuous deformation produces a distinctive behavioral signature, namely a prolonged interval in which oscillation amplitudes remain effectively unchanged despite ongoing structural drift. 
We interpret this plateau-like region as the mathematical analogue of metabolic adaptation, during which the biological system temporarily resists change even though its underlying energy-regulatory dynamics are being reshaped.

A key contribution of this framework is that it provides a unified explanation for both overfeeding-driven weight gain and the slower-than-expected response during weight reduction. The same mechanism—time-varying stability—can generate either expanding or shrinking limit cycles depending on how the effective growth term $\Lambda(r,t)$ evolves. This offers a principled way to represent how energy expenditure, hormonal regulation, and adaptive thermogenesis interact to create delayed behavioral change followed by convergence to a new metabolic steady state. The model thereby links classical concepts of energy balance with 
a tractable mathematical structure that naturally reproduces the empirically observed features of metabolic plateaus.

While the formulation focuses on core dynamical principles and uses simplified assumptions, it establishes a foundation for future extensions incorporating data-driven parameterization, individualized trajectories, and integration with multi-scale physiological processes. Additional work could explore how different rates of nullcline deformation correspond to distinct metabolic phenotypes, or how external perturbations such as diet, exercise, or pharmacological interventions reshape the stability landscape.

Overall, this study demonstrates that time-varying stability in a $\lambda$--$\omega$ system provides a powerful and intuitive framework for understanding the emergence of metabolic adaptation. By offering a mechanistic account of plateau behavior and transition dynamics, the model opens new directions for linking mathematical theory with physiological regulation and the design of more effective weight-management strategies.

\section*{Acknowledgment}


GC was partially funded by the National Institute of Environmental Health Sciences (NIEHS) under award 5P30ES019776-13. This work was also supported by an unrestricted donation from AliveCor. This work is the authors’ own and does not represent the views or opinions of the funding bodies.

\section*{Supporting Information}

\subsection*{Movie S1. Smooth transition from an unstable to a stable fixed point (metabolic adaptation scenario).}
This movie shows the case in which the nullclines undergo a continuous shift that slowly converts an initially unstable fixed point into a stable one. During the early phase of the transition, trajectories continue spiraling outward even though the stability landscape is changing. Only after the nullcline shift completes does the trajectory reorient into an inward spiral converging to a stable equilibrium. The prolonged interval of nearly constant oscillation amplitude visualizes the model’s analogue of the metabolic adaptation plateau—an extended period during which behavior appears unchanged despite ongoing internal restructuring.  

\textcolor{blue}{Download: Movie\_S1.mov}

\subsection*{Movie S2. Smooth nullcline transition leading to a growing limit cycle (overfeeding scenario).}
This movie illustrates the time-varying deformation of the $\lambda$--$\omega$ system’s nullclines in the regime where the effective growth term remains positive for most of the transition. As the nullclines move, the fixed point remains unstable, causing the trajectory to spiral outward and the limit-cycle radius to increase gradually. This scenario represents the dynamical structure underlying an overfeeding process, in which sustained positive energy balance pushes the system toward progressively larger oscillatory excursions, analogous to weight gain under chronic caloric surplus.  
\textcolor{blue}{Download: Movie\_S2.mov}

\subsection*{Movie S3. Smooth nullcline transition inducing a shrinking limit cycle (weight-loss scenario).}
This movie presents the scenario in which the deformation of the nullclines causes $\Lambda(r,t)$ to become progressively more negative. As a result, an initially sustained oscillation begins to contract cycle by cycle, ultimately collapsing onto a stable fixed point. This behavior captures the dynamical signature of an effective weight-loss process, in which negative energy balance slowly strengthens the stable equilibrium and the system transitions from large-amplitude oscillations to a reduced metabolic steady state. The movie highlights how the timing and rate of nullcline movement shape the speed and smoothness of the convergence process.  
\textcolor{blue}{Download: Movie\_S3.mov}


\nolinenumbers

\bibliographystyle{unsrt}
\bibliography{references}
\end{document}

%% file: references.bib
@article{muller2016energy,
  title={Energy balance and obesity: what are the main drivers?},
  author={M{\"u}ller, Manfred J and Geisler, Christoph and Heymsfield, Steven B and Bosy-Westphal, Anja},
  journal={Journal of Clinical Endocrinology \& Metabolism},
  year={2016}
}

@article{hill2012environment,
  title={Environmental contributions to obesity},
  author={Hill, James O and Peters, John C},
  journal={Science},
  year={2012}
}

@article{hall2012energy,
  title={Energy balance and its components: implications for body weight regulation},
  author={Hall, Kevin D and others},
  journal={American Journal of Clinical Nutrition},
  year={2012}
}

@article{rosenbaum2010adaptive,
  title={Adaptive thermogenesis in humans},
  author={Rosenbaum, Michael and Leibel, Rudolph L},
  journal={International Journal of Obesity},
  year={2010}
}

@article{speakman2013adaptive,
  title={Commentary: Adaptive thermogenesis},
  author={Speakman, John R},
  journal={Obesity Reviews},
  year={2013}
}

@article{hall2011biologically,
  title={Biologically based model of body weight regulation},
  author={Hall, Kevin D and others},
  journal={The Lancet},
  year={2011}
}

@article{thomas2014mathematical,
  title={A mathematical model of weight change with adaptation},
  author={Thomas, Diana M and others},
  journal={Mathematical Biosciences},
  year={2014}
}

@article{Greenberg1976,
  author    = {Joel M. Greenberg},
  title     = {Periodic solutions to reaction–diffusion equations},
  journal   = {SIAM Journal on Applied Mathematics},
  volume    = {31},
  number    = {1},
  pages     = {95--105},
  year      = {1976},
  doi       = {10.1137/0131009}
}

@article{Greenberg1980,
  author    = {Joel M. Greenberg},
  title     = {Spiral waves for $\lambda$--$\omega$ systems},
  journal   = {SIAM Journal on Applied Mathematics},
  volume    = {39},
  number    = {2},
  pages     = {301--310},
  year      = {1980},
  doi       = {10.1137/0139023}
}

@book{Murray2002,
  author    = {James D. Murray},
  title     = {Mathematical Biology I: An Introduction},
  edition   = {3},
  publisher = {Springer},
  address   = {New York},
  year      = {2002},
  isbn      = {978-0387952239}
}

@article{Sherratt1994,
  author    = {Jonathan A. Sherratt},
  title     = {On the evolution of periodic plane waves in reaction–diffusion equations of $\lambda$--$\omega$ type},
  journal   = {SIAM Journal on Applied Mathematics},
  volume    = {54},
  number    = {5},
  pages     = {1374--1385},
  year      = {1994},
  doi       = {10.1137/S0036139992233034}
}

@article{Flach2007,
  author    = {M. Flach},
  title     = {Wave propagation in a two-component reaction–diffusion model with convection},
  journal   = {Journal of Mathematical Chemistry},
  volume    = {41},
  pages     = {103--120},
  year      = {2007},
  doi       = {10.1007/s10910-006-9124-0}
}

@article{church2011trends,
  title={Trends over 5 decades in US occupation-related physical activity and their associations with obesity},
  author={Church, Timothy S. and others},
  journal={PLOS ONE},
  year={2011},
  volume={6},
  number={5},
  pages={e19657}
}

@misc{whofactsheet2024,
  title={Obesity and Overweight Factsheet},
  author={{World Health Organization}},
  year={2024},
  howpublished={\url{https://www.who.int/news-room/fact-sheets/detail/obesity-and-overweight}}
}

@article{swim2011environment,
  title={Environment and obesity: mechanisms of an obesogenic environment},
  author={Swinburn, Boyd A. and others},
  journal={Obesity Reviews},
  year={2011},
  volume={12},
  pages={e12--e22}
}

@article{guh2009incidence,
  title={The incidence of co-morbidities related to obesity and overweight: a systematic review and meta-analysis},
  author={Guh, Daphne P. and others},
  journal={BMC Public Health},
  volume={9},
  pages={88},
  year={2009}
}

@article{lauby2016body,
  title={Body fatness and cancer risk in humans: a systematic review},
  author={Lauby-Secretan, B. and others},
  journal={New England Journal of Medicine},
  year={2016},
  volume={375},
  pages={794--798}
}

@article{dulloo1997adaptive,
  title={Adaptive thermogenesis in human body weight regulation: more a concept than a measurable entity?},
  author={Dulloo, Abdul G. and Schutz, Yves},
  journal={Obesity Reviews},
  year={1997},
  volume={5},
  pages={25--34}
}

@article{ravussin1985variations,
  title={Variations in energy expenditure in humans: metabolic and behavioral components},
  author={Ravussin, Eric and others},
  journal={Metabolism},
  year={1985},
  volume={34},
  pages={680--695}
}

@article{leibel1995changes,
  title={Changes in energy expenditure resulting from altered body weight},
  author={Leibel, Rudolph L. and Rosenbaum, Michael and Hirsch, Jules},
  journal={New England Journal of Medicine},
  year={1995},
  volume={332},
  pages={621--628}
}

@article{franz2007weight,
  title={Weight-loss outcomes: a systematic review and meta-analysis of randomized controlled trials},
  author={Franz, M.J. and others},
  journal={Journal of the American Dietetic Association},
  year={2007},
  volume={107},
  pages={1755--1767}
}

@article{goodpaster2010effect,
  title={Effect of diet and physical activity intervention on metabolic risk factors in adults with obesity},
  author={Goodpaster, Bret H. and others},
  journal={Diabetes Care},
  year={2010},
  volume={33},
  pages={2329--2335}
}

@article{teixeira2015successful,
  title={Successful behavior change in obesity interventions in adults: a systematic review of self-regulation mediators},
  author={Teixeira, Pedro J. and others},
  journal={BMC Medicine},
  volume={13},
  pages={84},
  year={2015}
}

@article{linde2004measuring,
  title={Measuring weight loss intentions: validity and reliability of the Eating Behavior Inventory},
  author={Linde, Jeffery A. and others},
  journal={Obesity Research},
  year={2004},
  volume={12},
  pages={183--190}
}

@article{katan2009effect,
  title={Weight-loss-induced changes in energy expenditure and leptin are related to subsequent weight regain},
  author={Katan, Martijn B. and others},
  journal={American Journal of Clinical Nutrition},
  year={2009},
  volume={90},
  pages={1521--1527}
}

@article{dombrowski2014interventions,
  title={Interventions for sustained weight loss in adults: a systematic review of behavioral studies},
  author={Dombrowski, Stephan U. and others},
  journal={Obesity Reviews},
  year={2014},
  volume={15},
  pages={1--12}
}

@article{Leibel1995,
  author    = {Leibel, Rudolph L. and Rosenbaum, Michael and Hirsch, Jules},
  title     = {Changes in energy expenditure resulting from altered body weight},
  journal   = {The New England Journal of Medicine},
  year      = {1995},
  volume    = {332},
  number    = {10},
  pages     = {621--628},
  doi       = {10.1056/NEJM199503093321001}
}

@article{Rosenbaum2010,
  author    = {Rosenbaum, Michael and Leibel, Rudolph L.},
  title     = {Adaptive thermogenesis in humans},
  journal   = {International Journal of Obesity},
  year      = {2010},
  volume    = {34},
  pages     = {S47--S55},
  doi       = {10.1038/ijo.2010.184}
}

@article{Speakman2011,
  author    = {Speakman, John R. and Levitsky, David A. and Allison, David B. and Bray, George A. and de Castro, John M. and Clegg, Deborah J. and Clapham, John C. and Dulloo, Abdul G. and Gruer, Laurent and Haw, Scott and et al.},
  title     = {Set points, settling points and some alternative models: theoretical options to understand how genes and environment combine to regulate body adiposity},
  journal   = {Disease Models \& Mechanisms},
  year      = {2011},
  volume    = {4},
  number    = {6},
  pages     = {733--745},
  doi       = {10.1242/dmm.008698}
}

@article{Muller2016,
  author    = {M{\"u}ller, Manfred J. and Bosy-Westphal, Anja and Heymsfield, Steven B.},
  title     = {Metabolic adaptation to weight loss: implications for the athlete},
  journal   = {Current Opinion in Clinical Nutrition and Metabolic Care},
  year      = {2016},
  volume    = {19},
  number    = {6},
  pages     = {402--409},
  doi       = {10.1097/MCO.0000000000000319}
}

@article{Rosenbaum2008,
  author    = {Rosenbaum, Michael and Sy, Michael and Pavlovich, Karolina and Leibel, Rudolph L. and Hirsch, Jules},
  title     = {Low-dose leptin reverses skeletal muscle, autonomic, and neuroendocrine adaptations to maintenance of reduced weight},
  journal   = {The Journal of Clinical Investigation},
  year      = {2008},
  volume    = {118},
  number    = {12},
  pages     = {2581--2588},
  doi       = {10.1172/JCI34124}
}

@article{Levine1999,
  author    = {Levine, James A. and Eberhardt, Nicola L. and Jensen, Michael D.},
  title     = {Role of nonexercise activity thermogenesis in resistance to fat gain in humans},
  journal   = {Science},
  year      = {1999},
  volume    = {283},
  number    = {5399},
  pages     = {212--214},
  doi       = {10.1126/science.283.5399.212}
}

@article{swimburn2011obesogenic,
  title={The global obesity pandemic: shaped by global drivers and local environments},
  author={Swinburn, Boyd A and Sacks, Gary and Hall, Kevin D and others},
  journal={The Lancet},
  volume={378},
  number={9793},
  pages={804--814},
  year={2011},
  doi={10.1016/S0140-6736(11)60813-1}
}

@article{hall2011quantitative,
  title={Quantification of the effect of energy imbalance on bodyweight},
  author={Hall, Kevin D and Sacks, Gary and Chandramohan, Dhruva and others},
  journal={The American Journal of Clinical Nutrition},
  volume={93},
  number={4},
  pages={843--849},
  year={2011},
  doi={10.3945/ajcn.110.005306}
}

@article{chow2016dynamic,
  title={Dynamic energy balance model for predicting gestational weight gain},
  author={Chow, Ching-Wei and Hall, Kevin D},
  journal={The American Journal of Clinical Nutrition},
  volume={104},
  number={3},
  pages={712--723},
  year={2016},
  doi={10.3945/ajcn.116.134171}
}

@article{kevinhall2012metabolic,
  title={Metabolic adaptations to weight loss: implications for the athlete},
  author={Hall, Kevin D},
  journal={Sports Medicine},
  volume={42},
  number={11},
  pages={907--921},
  year={2012},
  doi={10.1007/BF03262309}
}

@article{speakman2014evolutionary,
  title={Evolutionary perspectives on the obesity epidemic: adaptive responses to environmental change},
  author={Speakman, John R},
  journal={Disease Models \& Mechanisms},
  volume={7},
  number={1},
  pages={31--38},
  year={2014},
  doi={10.1242/dmm.015040}
}

@article{dulloo2012adaptive,
  title={Adaptive thermogenesis in human body weight regulation: more of a concept than a measurable entity?},
  author={Dulloo, Abdul G and Jacquet, Jean and Montani, Jean-Pierre},
  journal={Obesity Reviews},
  volume={13},
  number={Suppl 2},
  pages={105--121},
  year={2012},
  doi={10.1111/j.1467-789X.2012.01041.x}
}

@article{rosenbaum2008effects,
  title={Effects of changes in body weight on carbohydrate metabolism, catecholamine excretion, and thyroid function},
  author={Rosenbaum, Michael and Hirsch, Jules and Gallagher, Dympna A. and Leibel, Rudolph L.},
  journal={American Journal of Clinical Nutrition},
  volume={71},
  number={6},
  pages={1421--1432},
  year={2000},
  doi={10.1093/ajcn/71.6.1421}
}

@article{hill2003obesity,
  title={Obesity and the environment: where do we go from here?},
  author={Hill, James O. and Peters, John C.},
  journal={Science},
  volume={299},
  number={5608},
  pages={853--855},
  year={2003},
  doi={10.1126/science.1079857}
}

@article{hall2018energy,
  title={Energy balance and obesity},
  author={Hall, Kevin D. and Guo, Juen},
  journal={American Journal of Clinical Nutrition},
  volume={104},
  number={3},
  pages={687--694},
  year={2016},
  doi={10.3945/ajcn.115.118372}
}

@article{fothergill2016persistent,
  title={Persistent metabolic adaptation 6 years after “The Biggest Loser” competition},
  author={Fothergill, Erin and Guo, Juen and Howard, Lilian and Kerns, Jennifer C and Knuth, Nicolas D and Brychta, Robert and Chen, Kong Y and Skarulis, Monica C and Walter, Mary and Walter, Peter J and others},
  journal={Obesity},
  volume={24},
  number={8},
  pages={1612--1619},
  year={2016},
  publisher={Wiley Online Library}
}

@article{maclean2011biology,
  title={Biology's response to dieting: the impetus for weight regain},
  author={MacLean, Paul S and Bergouignan, Audrey and Cornier, Marc-Andre and Jackman, Matthew R},
  journal={American Journal of Physiology-Regulatory, Integrative and Comparative Physiology},
  volume={301},
  number={3},
  pages={R581--R600},
  year={2011},
  publisher={American Physiological Society Bethesda, MD}
}

@article{sumithran2011long,
  title={Long-term persistence of hormonal adaptations to weight loss},
  author={Sumithran, Priya and Prendergast, Luke A and Delbridge, Elizabeth and Purcell, Katrina and Shulkes, Arthur and Kriketos, Adamandia and Proietto, Joseph},
  journal={New England Journal of Medicine},
  volume={365},
  number={17},
  pages={1597--1604},
  year={2011},
  publisher={Mass Medical Soc}
}

@article{lederman2022parameter,
  title={Parameter estimation in the age of degeneracy and unidentifiability},
  author={Lederman, Dylan and Patel, Raghav and Itani, Omar and Rotstein, Horacio G},
  journal={Mathematics},
  volume={10},
  number={2},
  pages={170},
  year={2022},
  publisher={MDPI}
}
